\documentstyle{article}

\author{Tofig M.Gassym\thanks{The results of this paper was reported on the V International Conference''Tousi-800 School''Shamakhy Astrophysical Observatory 14-17 October,1999 and the thesis of report is printed in [17]. }\\
Institute of Physics Academy of Sciences of Azerbaijan Republic, \\
Baku-370143,H.Javid Aven.33.}
\title{The interacting system of electrons, positrons and photons in high external electric and arbitrary magnetic fields
}
\input tcilatex

\QQQ{Language}{
American English
}

\begin{document}

\maketitle
\begin{abstract}
The connected system of Boltzman equations for the interacting system of
electrons, positrons and photons in high external electric E and arbitrary
magnetic H fields is solved. The consideration is made under the conditions
of arbitrary heating and the mutual drag of carriers and photons.

The non-stationary and non-uniform distribution function (DF) of photons for
the all considered cases is obtained. It is shown that the DF of photons
have the stationary limit for the drift velocities $\left( \stackrel{%
\rightarrow }{u}\stackrel{\rightarrow }{q}/\hbar \omega _q\right) <1$ and
becomes exponentially grown by the time for the drift velocities $\left( 
\stackrel{\rightarrow }{u}\stackrel{\rightarrow }{q}/\hbar \omega _q\right)
\geq 1$. It is shown that the mutual drag of carriers and photons leads to
the formation of ''quasi-particles''\_ the ''electron dressed by photons''
and ''positron dressed by photons'', i.e. the mutual drag plays the role of
the dressing mechanism of carriers and leads to the renormalization of the
mass and frequency of photons.

As a result of analyses of the phenomena of connected by the mutual drag
system of carriers and photons we receive some fundamental results: a) the
finiteness of the mass of photon (i.e. the rest mass of photons do not equal
to zero); b) reality of takhyons as a quasi-particles with the real mass in
amplifying system (or regime); c) identity of the mechanism of relativity
and the Doppler effect which coincides with the renormalization of frequency
or mass as a result of the mutual drag of carriers and photons at external
field (force). These conclusions were received as a result of the fact that
the relativistic factor enters to the expressions of the DF of photons and
other physical expressions in the first order in the form $\left[ 1-\left(
u^2/c^2\right) \right] ^{-1}$, instead the

$\left[ 1-\left( u^2/c^2\right) \right] ^{-1/2}$ in Einstein theory. It is
shown that the relativistic effect of deceleration of time really taking
place for the relaxation time of carriers or for the life time of carriers,
or for the period of electromagnetic oscillations. The latest is a direct
result of the Doppler effect. Also it is shown that the velocity of light is
an averaged velocity of photons at the ground state, as well as a velocity
of sound for the phonons.
\end{abstract}

\section{Introduction}

In earlier my publications made the theoretical investigation for the
interacting system of electrons and phonons in semiconductors, semimetals
and gaseous plasmas at high external electric and magnetic fields also under
the conditions of the propagation of strong electromagnetic waves. In [1-4]
the connected system of kinetic equations of interacting electrons, holes
and phonons at high electric and magnetic fields was solved, with taking
into account arbitrary heating and the mutual drag of carriersand phonons.
For the phonons the solution of non-stationary kinetic equation was found
and was shown that the non-equilibrium and non-stationary distribution
function of phonons has the stationary limit, when the drift velocity of
connected by the mutual drag system of carriers and phonons is less than the
velocity of sound $(u<s)$. For the drift velocities $u>s$ the distribution
function of phonons becomes is exponentially grown by the time. It accords
to the generation of intrinsic phonons and amplification of phonons
introduced to the system externally (the stream of phonons).

It was shown that the mutual drag leads to the re-normalization the mass of
carriers. As a result of the mutual drag electrons and holes are ''dressed''
by phonons and formed the ''quasi-particles'' which has the electrons or
holes charge $(\pm e)$ and the phonons mass $m=T_i/s^2$ (where $T_i$ - is
the temperature of the coupled by the mutual drag system of carriers and
phonons) [1-5]. In weak electric fields the mass of phonons are $m_o=T/s^2$.
In the case of propagation of strong electromagnetic wave in semiconductors
and semimetals at external magnetic field this lead to the cyclotron
resonance on such ''quasi-particles'' with the frequencies $\omega _H$$%
=(3/4)eH/\left[ (T_i/s^2)\cdot c\right] $ [6,7]. It was shown in [7], that
the effect of connection carriers and phonons as a result of their mutual
drag is a common phenomenon. Really, as it follows from [7] under the
conditions of longitudinal propagation of strong electromagnetic wave in
semimetals with equal electron and hole concentrations the frequencies of
Alfven and magneto-sound waves are identically equal to each other and have
the form:

$$
\omega _A=\omega _{ms}=kH/\left( 4mN\right) ^{1/2}=kv_A, 
$$
where $m_i=4(T_i/s^2)/3$ also is so-called '' mass of phonons'', $%
T_i=T_i(+0) $ - the effective temperature of the connected by the mutual
drag system of ''electron + phonons'' or ''hole + phonons''. Thus we may
conclude that the mutual drag of carriers and phonons at external fields
lead to the formation of compound particles (''quasi-particles'') -
''carriers dressed by phonons'' with the joint drift velocity u.

In the first investigation [9] the considerations was made for the
degenerate semiconductors and semimetals under the conditions of the
propagation of weak electromagnetic waves. In particular in [8] was
considered the propagation of weak electromagnetic waves on degenerate
semiconductors with one type of carriers and was predicted the cyclotron
resonance on the connected system of electron + phonon. But soon it was
shown in [10] that this resonance does not observable because of the
presence of impurities with the concentrations $N_i=n$ (where $n$ is the
concentration of electrons). Also in [10] was shown that in pure semimetals
it is necessary to take into account the presence of two type of carriers
which drifted on the opposite directions at $H=0$. The remarks about
non-observably in [10] applies also to the uniform and non-uniform low
frequency cyclotron resonance in degenerate semiconductors and semimetals
with one type of carriers under the conditions of propagation strong
electromagnetic waves considered in [6,7] and [11]. These questions was
discussed in [3,7] and was shown that in intrinsic semiconductors and
semimetals realized both uniform and non-uniform cyclotron resonance on
quasi-particle ''hole + phonon''. In [3,7] also was shown that the such type
of resonance realized on non-degenerate impurity and intrinsic
semiconductors and discussed the question about their observably.

\section{General}

In present paper the connected system of kinetic equations for interacting
system of electrons, positrons and photons in external high electric $%
\stackrel{\rightarrow }{E}$ and arbitrary magnetic $\stackrel{\rightarrow }{H%
}$ fields are solved. The non-equilibrium distribution functions of
electrons, positrons and photons are founded by the taking into account of
the arbitrary heating and mutual drag.

For the photons the general solution of non-stationary and non-uniform
Boltzmann equations was founded. The cases of week $(\omega _H^{\pm }\tau
_c)\ll 1$ , classically high $(\omega _H^{\pm }\tau _c)\gg 1$, and
quantizing $(\hbar \omega _H^{\pm }\gg T,T_c)$ high magnetic fields are
considered. Here $T$- is the initial temperature of equilibrium system (at
the $t=0$, before the external fields is applied), $T_c(\stackrel{%
\rightarrow }{E},\stackrel{\rightarrow }{H})$ -is the temperature of heated
electrons and positrons, $\omega _H^{\pm }$ - is their cyclotron
frequencies, $\tau _c^{-1}=\nu _c$ the relaxation frequencies of electrons
and positrons( the photon production by the annihilation electron-positron's
pair or by the scattering of electrons and positrons by photons). $c=e,p$
means electrons and positrons, accordingly.

In the absence of or in a weak magnetic fields the inter-carrier collisions
frequencies $\nu _{ee}$ and $\nu _{pp}$ assumed much more than others, and
that is why the isotropic parts of the distribution functions of carriers
assumed to be equilibrium one, with effective temperatures of carriers $%
T_c=T_{e,p}(E,H)$. This approximation corresponds to the case of high
concentration of carriers $n>n_k$.

If the external field is a field of strong electromagnetic wave, then it is
necessary to fulfil of the condition:

\begin{equation}
\label{1}\omega \gg \nu _\epsilon ^{\pm } 
\end{equation}
$\nu _\epsilon ^{\pm }$- is the collusion frequencies of carriers for energy
transfer to scatters, $\omega $ - is the frequency of electromagnetic wave.

Under the conditions (1) the isotropic parts of carriers distribution
functions at zero approximation by $\nu _\epsilon ^{\pm }/\omega $ do not
depend on time directly. By the direct solution of quantum kinetic equations
in common case for the arbitrary spherical symmetric dispersion low of
carriers, it was shown that at quantizing and classically high magnetic
fields the stationary distribution functions of carriers, satisfying the
boundary conditions, $F^{\pm }(\epsilon )\mid _{\epsilon \rightarrow \infty
}=0$ has the form (for the arbitrary quantities of their concentrations):

\begin{equation}
\label{2}F^{\pm }(\epsilon )=\left\{ c^{-1}\exp \left( \int d\epsilon
^{`}/T_c\left( \epsilon ^{`},t\right) +1\right) \right\} ^{-1} 
\end{equation}
There $T_c(\epsilon ,t)=A(\epsilon ,t)/B(\epsilon )$ - the temperature of
carriers which have occupied the energetic level $\epsilon $,

$$
A(\epsilon )=(2\pi /\hbar )\sum_{\alpha \beta q}C_q^2\cdot \mid I_{\alpha
\beta }^2\mid \cdot \left( \hbar \omega _q^{*}\right) ^2N\left( q,t\right)
\delta \left( \epsilon _\beta -\epsilon _\alpha -\hbar \omega _q\right)
\cdot \delta \left( \epsilon _\alpha -\epsilon \right) 
$$

$$
B(\epsilon )=(2\pi /\hbar )\sum_{\alpha \beta q}C_q^2\cdot \mid I_{\alpha
\beta }^2\mid \cdot \hbar \omega _q\cdot \delta \left( \epsilon _\beta
-\epsilon _\alpha -\hbar \omega _q\right) \cdot \delta \left( \epsilon
_\alpha -\epsilon \right) \text{, }\omega _H^{*}=\hbar \omega _q-\stackrel{%
\rightarrow }{V}\stackrel{\rightarrow }{q} 
$$
$C_q$- being the constant of interaction and $I_{\alpha \beta }$ - is a
matrix elements for the transition from state $\alpha $ to $\beta $ and back
(reverse).

For the arbitrary degree of quantization we have:

$$
F^{\pm }(\epsilon )=\left\{ 1+\exp \left[ (\epsilon -\zeta \left( E,H\right)
/T_c\right] \right\} ^{-1} 
$$

\begin{equation}
\label{3}T_c=T_i\left\{ 1+(\frac{V^{\pm }}u-)-1]2(f1-1)\}\right\} 
\end{equation}

$$
\varphi _1=\left[ 1-\frac{u^2}{c^2}\right] ^{-1/2} 
$$
At the classical region of strong magnetic field we have:

$$
T_c=T_i\left\{ 1+\frac 13\left( \frac{V^{\pm }}c\right) ^2+\left[ 1-\frac{%
V^{\pm }}u\right] \left( \varphi _2-1\right) \right\} ,\text{ }\varphi
_2=\frac c{2u}\ln \left| \frac{c+u}{c-u}\right| \text{ (3')} 
$$
Here $V^{\pm }=cE/H$ is the Hall's drift velocity of carriers. It was shown
that for all considered cases the solution of non-stationary kinetic
equation for the photons is:

$$
N(\stackrel{\rightarrow }{q},t)=\left\{ N(\stackrel{\rightarrow }{q},%
\stackrel{\rightarrow }{r}-\stackrel{\rightarrow }{u_o}t,0)+\beta
\int\limits_0^tN(q,\tau `)exp\left( -\int\limits_0^\tau \gamma _q\left( \tau
`\right) d\tau `\right) \right\} \times 
$$

\begin{equation}
\label{4}\times exp\left[ \int\limits_0^t\gamma _q\left( \tau \right) d\tau
\right] 
\end{equation}

Where $N(\stackrel{\rightarrow }{q},\stackrel{\rightarrow }{r}-\stackrel{%
\rightarrow }{u_o}t,0)$ the distribution function of photons in the absence
of electric and magnetic fields (at $t=0$), which in the case of space
uniformity is equilibrium Plank's function at the temperature $T$. The
increasing increment of photons is

\begin{equation}
\label{5}\gamma _q=\beta \left[ \frac{\stackrel{\rightarrow }{u}\stackrel{%
\rightarrow }{q}}{\hbar \omega _q}-1\right] 
\end{equation}
$\beta =\left( \beta _e+\beta _p+\beta _{ph}+\beta \right) $is a total
collisions frequency of photons with electrons $(e)$, positrons $(p)$
(including the photon decay to electron-positron pair), photons $(ph)$ and
boundaries $(b)$ of region occupied by the system, if such one exists.

\begin{equation}
\label{6}\stackrel{\rightarrow }{u}(t)=\sum\limits_{\pm }\stackrel{%
\rightarrow }{u^{\pm }}(t)=\frac{\beta _e}\beta \stackrel{\rightarrow }{V^-}%
(t)+\frac{\beta _p}\beta \stackrel{\rightarrow }{V^{+}}(t) 
\end{equation}
$\stackrel{\rightarrow }{V^{\pm }}(t)$ is average drift velocity of
carriers, $\stackrel{\rightarrow }{u^{-}}(t)$ - the drift velocity of
connected by the mutual drag system of ''electron + positrons'' and $%
\stackrel{\rightarrow }{u^{+}}(t)$ is the same for the ''positron +
photons''.

In the common case, when the heating of carriers was realized by the field
of strong electromagnetic wave $\stackrel{\rightarrow }{E}=\stackrel{%
\rightarrow }{E_o}e^{-i\omega t}+\stackrel{\rightarrow }{E_o^{*}}e^{+i\omega
t}$, $\stackrel{\rightarrow }{V^{\pm }}(t)=\stackrel{\rightarrow }{V^{\pm }}%
\cos \omega t$ we have:

$$
N(\stackrel{\rightarrow }{q},t)=\left\{ N(q,o)+\beta \int\limits_0^td\tau
N(q,\tau )\exp \left[ \beta \left( \tau -\frac{\stackrel{\rightarrow }{u}%
\stackrel{\rightarrow }{q}}{\hbar \omega _q}\frac{\sin \omega \tau }\omega
\right) \right] \right\} \times 
$$

\begin{equation}
\label{7}\times exp\left\{ -\beta \left[ t-\frac{\stackrel{\rightarrow }{u}%
\stackrel{\rightarrow }{q}}{\hbar \omega _q}\frac{\sin \omega t}\omega
\right] \right\} 
\end{equation}
In the case of constant external electric field $(\omega \rightarrow 0)$ we
have:

\begin{equation}
\label{8}N(\stackrel{\rightarrow }{q},t)=\left\{ \frac{N(q,o)-N(q,T_i)}{1-%
\stackrel{\rightarrow }{u}\stackrel{\rightarrow }{q}/\hbar \omega _q}%
\right\} exp\left\{ \beta \left[ \frac{\stackrel{\rightarrow }{u}\stackrel{%
\rightarrow }{q}}{\hbar \omega _q}-1\right] t\right\} +\frac{N(q,T_i)}{1-%
\stackrel{\rightarrow }{u}\stackrel{\rightarrow }{q}/\hbar \omega _q} 
\end{equation}

Here $T_i=\left( \beta _c/\beta \right) T_c+\left( \beta _{ph}/\beta \right)
T_{ph}+\left( \beta _b/\beta \right) T_b$ is the temperature of the coupled
by the mutual drag system of heated complexes of carriers and photons.

We have still considered the case when the initial state of photons (at $t=0$%
) was assumed to be equilibrium state without distinguished direction. If
the part of initial distribution of photons has directional drift (the
photons stream), then the kinetic equation for the photons has the form:

\begin{equation}
\label{9}\frac{\partial N(q,\stackrel{\rightarrow }{r},t)}{\partial t}+\frac{%
\partial N(q,\stackrel{\rightarrow }{r},t)}{\partial \stackrel{\rightarrow }{%
r}}\frac{d\stackrel{\rightarrow }{r}}{dt}=\beta \left\{ -N(q,t)\left( 1- 
\frac{\stackrel{\rightarrow }{u}\stackrel{\rightarrow }{q}}{\hbar \omega _q}%
\right) \right\} +N_i\left( q,T_i\right) 
\end{equation}

By the single substitution $t`=t$ and $\stackrel{\rightarrow }{r`}=\stackrel{%
\rightarrow }{r}-\stackrel{\rightarrow }{u_o}t$ the solution (10) may be
transformed into the form:

$$
N(\stackrel{\rightarrow }{q},\stackrel{\rightarrow }{r},t)=\left\{ N(%
\stackrel{\rightarrow }{q},\stackrel{\rightarrow }{r}-\stackrel{\rightarrow 
}{u_o}t,0)+\beta N_i(q,T_i)\int\limits_0^texp\left[ \beta \int\limits_0^\tau
\left( 1-\frac{\stackrel{\rightarrow }{u}\stackrel{\rightarrow }{q}}{\hbar
\omega _q}\right) d\tau `\right] d\tau \right\} \times 
$$

$$
\times exp\left[ -\beta \int\limits_0^\tau \left( 1-\frac{\stackrel{%
\rightarrow }{u}\stackrel{\rightarrow }{q}}{\hbar \omega _q}\right) d\tau
\right] \text{ (9')} 
$$

Thus it is seemed that the solution of the uniform equation for the photons
(8) and non-uniform equation (9') have the same form corresponding to the
different initial conditions. Therefore, if the initial distribution
function of photons has the form of the drifted Plank distribution function $%
N(q,\stackrel{\rightarrow }{r}-\stackrel{\rightarrow }{u_o}t,0)$ and if the
external fields is uniform, then this non-uniformity has to be served with
time and the drift at external field is to be added to them. Thus the
equation (9') allows to consider processes of absorption or amplification of
photons, introduced to the system from outside (initial stream of photons),
and the generation of own photons of system in external fields. In principle
it is the most common form of the initial distribution function, which is
taking the chance for examination of the affirmation of the special theory
of relativity about equivalency of all inertial frames of reference. Really
by the transition to the frame of reference drifting jointly with photons,
as a result, we have the Plank's equilibrium distribution function at the
temperature $T$ in this frame of reference. In the other words, in the
absence of external fields $(E=H=0)$ for the initial non-uniform system of
photons (9) from kinetic equation we receive the uniform one, by the
transition from one frame of reference to an other but it is not means that
the two frames of reference is equivalent.

In fact, by the transition from the frame of reference drifting jointly with
the photons by the velocity$\stackrel{\rightarrow }{u_o}=c\stackrel{%
\rightarrow }{q}/q$ to the frame of reference which in the rest which is
equivalent to the transition from one internal frames of reference $(u=u_o)$
to the other $(u=0)$ we receive $N(\stackrel{\rightarrow }{q},\stackrel{%
\rightarrow }{r}-\stackrel{\rightarrow }{u_o}t,0)=N(q,0)=N_o(q,T)$, for all
moments of time, that is the solution (9') transform to (8), but it is not
means that this two frame of reference is equivalent. On the other words the
demand of equivalency of the laws of physics in that two inertial frame of
reference is equivalent to the demand of the equivalency the equilibrium
Plank's distribution function to the drifted Plank's distribution function
or to the demand of equivalency the laws of physics in the uniform and
non-uniform cases (or spaces).

As one can see from (4), (8) and (9') the general solution of non-stationary
equation of photons have the stationary limit in the region of drift
velocities $\left( \stackrel{\rightarrow }{u}\stackrel{\rightarrow }{q}%
/\hbar \omega _q\right) <1$

\begin{equation}
\label{10}\lim _{t\rightarrow \infty }N(\stackrel{\rightarrow }{q},t)=N(%
\stackrel{\rightarrow }{q})=N(q,Ti)\left( 1-\frac{\stackrel{\rightarrow }{u}%
\stackrel{\rightarrow }{q}}{\hbar \omega _q}\right) 
\end{equation}
As it seems from equations (4), (8) and (9') for the drift velocities $%
\left( \stackrel{\rightarrow }{u}\stackrel{\rightarrow }{q}/\hbar \omega
_q\right) >1$ the distribution function of photons becomes exponentially
grown by the time. It is accords to the generation of intrinsic photons by
the increment of grow $\gamma _q$ and amplification of photons introduced to
the system from outside (stream of photons) by the coefficient of
amplification

\begin{equation}
\label{11}\Gamma _q=\frac{\gamma _q}c=\frac \beta c\left( \frac{\stackrel{%
\rightarrow }{u}\stackrel{\rightarrow }{q}}{\hbar \omega _q}-1\right) =\frac
\beta c\left[ \frac uc\cos a-1\right] 
\end{equation}

$\alpha $$=\left( \stackrel{\rightarrow }{u}\symbol{94} \stackrel{%
\rightarrow }{q}\right) $ - the angle between the drift velocity of
connected by the mutual drag system ''electron + photons'' (''the dressed
electron'') or '' positron + photons'' (''the dressed positron'') and
momentum of photon. The expression for the electrical current of electrons
and positrons at classically high external magnetic fields has the form:

In the case of the propagation of strong electromagnetic wave and at the
presence of external magnetic field the current of electrons and positrons
has the form:

$\stackrel{\rightarrow }{j_{\pm }}=ne\stackrel{\rightarrow }{V^{\pm }},$ $%
\stackrel{\rightarrow }{V^{\pm }}=\left\langle \stackrel{\rightarrow }{%
V^{\pm }}(\epsilon )\right\rangle $ - is the averaged drift velocity of
carriers. Here

\begin{equation}
\label{12}\stackrel{\rightarrow }{V^{\pm }}\left( \epsilon \right) =\mp 
\frac{e\Omega _{\pm }(\epsilon )}{m_c}\cdot \frac{\stackrel{\rightarrow }{%
E_{\mp }}\left( \omega _H^{\pm }/\Omega _{\pm }\left( \epsilon \right)
\right) \left[ \stackrel{\rightarrow }{h}\stackrel{\rightarrow }{E}\right]
-\left( \omega _H^{\pm }/\Omega _{\pm }\left( \epsilon \right) \right) ^2%
\stackrel{\rightarrow }{h}\left( \stackrel{\rightarrow }{h}\stackrel{%
\rightarrow }{E}\right) }{\Omega _{\pm }^2-\left( \omega _H^{\pm }\right) ^2}%
,\text{ }\stackrel{\rightarrow }{h}=\stackrel{\rightarrow }{H}/H\text{ } 
\end{equation}

In the case of $\omega \rightarrow 0,$ $\stackrel{\rightarrow }{E}\Vert 
\stackrel{\rightarrow }{H}$ (or $H=0$):

\begin{equation}
\label{13}\stackrel{\rightarrow }{V^{\pm }}\left( \left\langle \epsilon
\right\rangle \right) =\frac{e\stackrel{\rightarrow }{E}\cdot \beta _c}{%
m_c\nu _{ph}\left( \left\langle \epsilon \right\rangle ,u\right) \cdot \beta
_{ph,b}}=\frac{e\stackrel{\rightarrow }{E}}{m\left( T_i,u\right) \cdot \beta
_{ph,b}} 
\end{equation}
Here $m=m_c\cdot \nu _{ph}\left( \left\langle \epsilon \right\rangle
,u\right) /\beta _c$-the mass of connected by the mutual drag system of
carriers and photons and $\beta _{ph,b}=\beta _{ph}+\beta _b$.

Really the interacting system of electrons, positrons and photons at the
external high electromagnetic and the classically high or the quantizing
magnetic fields under the conditions of their heating, mutual drags and at
the stationary conditions $\left( \stackrel{\rightarrow }{u}\stackrel{%
\rightarrow }{q}/\hbar \omega _q\right) <1$ has the cyclotron resonance with
the frequencies

\begin{equation}
\label{14}\omega _H^{\pm }=\frac{eH}{\left( T_i/c^2\right) c}=\frac{eH}{%
m(T_i)c} 
\end{equation}

As it seems from the equation (12) the resonance is taking place on
frequencies of electromagnetic wave less than the collision frequencies of
photons with carriers. The width of the resonance lines defined by the
expression $\gamma $$=(3/2)\left[ \omega ^2/\beta _c+\beta _{ph}+\beta
_b\right] $.

In the other words, because of the mutual drag, the electrons and positrons
turn into the compound particles (into the coupled system of ''electron +
photons'' and ''positrons + photons'' i.e. so called ''dressed'' by the
photons ''quasi-electron'' or ''quasi-positron'' with the effective mass $%
m(T_i)$. In fact, we receive the quasi-particle with the electron's or
positrons charge and the photons mass:

\begin{equation}
\label{15}m(T_i)=m(E,H)=T_i/c^2\text{ or }E=T_i=m(T_i)c^2 
\end{equation}

Since $T_i$ and $T$ means the average kinetic energy, then from the equation
(10), (11) and (14) we receive that the, so-called ''velocity of light'' in
vacuum $"c"$ is the average velocity of photons in the equilibrium or
stationary state with the temperatures $T$. The ''dressing''of electrons and
positrons in quantum electrodynamics was connected with the virtual
absorption and emission of photons by the electrons and positrons, which is
occupied the given level of energy, i.e. with the finiteness of the lifetime
of electrons and positrons or the natural width of the given energetic
level. For the interacting system of electrons, holes and phonons in
semiconductors, semimetals and gaseous plasmas the analogous problem was
solved in [1- 5].

As it seems from (4), (8), (9') and (10) at the external electric and
magnetic fields under the stationary conditions the relativistic factor
enters to the distribution function of photons in first order in the form $%
[1-(\stackrel{\rightarrow }{u}\stackrel{\rightarrow }{q}/\hbar \omega
_q)]^{-1}$, instead of the form $[1-(v^2/c^2)]^{-1/2}$ in relativistic
electrodynamics. This is connected with the violation of $T$- symmetry $%
(t\rightarrow -t)$ of equations in electrodynamics and, in common dynamics
at the external fields. In this case the uniformity of the space and as a
result the law of the conservation of momentum is violated too. Since the
external fields acts constantly but not instantaneously i.e. we have the
motion with acceleration and the oscillatory regime is absent. The
substitution $t\rightarrow -t$ do not simply lead to substitution $\stackrel{%
\rightarrow }{v}\rightarrow -\stackrel{\rightarrow }{v}$, because of the
motions along and opposite of field direction are differs and do not
compensate each other.

The relativistic factor of the type $[1-(v^2/c^2)]^{-1}$ may appears only at
the absence of external field in equilibrium or stationary conditions, by
the using the isotropic part of the distribution function of photons.
Factually, by the separation of thestationary distribution function of
photons (10) to the isotropic and anisotropy parts we have:

\begin{equation}
\label{16}N(q)=N_s(q)+N_\alpha (q)=N(q,T_i)\left[ 1-\frac{u^2}{c^2}\cos
{}^2a\right] ^{-1}+N(q,T_i)\frac uc\cos \alpha \left[ 1-\frac{u^2}{c^2}\cos
{}^2a\right] ^{-1} 
\end{equation}

Since as a result of the mutual drag the carriers and photons form the
connected system (complex) with the common drift velocity, then under the
conditions of strong (full) mutual drag $\alpha =0$ or $\pi $ and we
received:

\begin{equation}
\label{17}N_s(q)=N\left( q,T_i\right) \left[ 1-\frac{u^2}{c^2}\right] -1; 
\text{ }N_a(q)=\left( \frac uc\cos \alpha \right) N\left( q,T_i\right)
\left[ 1-\frac{u^2}{c^2}\right] ^{-1} 
\end{equation}
In the absence of external electric and magnetic fields, in common case $%
u=u_o=const$ and we have:

\begin{equation}
\label{18}N(\stackrel{\rightarrow }{q},\stackrel{\rightarrow }{u_o})=N(%
\stackrel{\rightarrow }{q},\stackrel{\rightarrow }{r}-\stackrel{\rightarrow 
}{u_o}t,0)=\left\{ exp\left( \frac{\hbar \omega _q^{*}}T\right) -1\right\}
^{-1} 
\end{equation}
$\hbar \omega _q$$=\hbar \omega _q-\stackrel{\rightarrow }{u_o}\stackrel{%
\rightarrow }{q}$. By the transition to the frame of reference drifting
together with photons we can received (16) the equilibrium Plank's
distribution function with the temperature $T$. Let us discuss the main
question now: may the presence of the relativistic factor of first or
secondary order lead to any singularities in physical phenomena or
quantities? As it seems from the non-stationary solution for the
distribution function of photons (8) or (9') the Lorenz-Einstein theory
corresponds to uniform (equilibrium) case and must satisfy the stationary
condition $v/c<1!$ The case of $v=c$ is not included to their theory and for
this reason the conclusions of the Einstein theory about equality the rest
mass of photons to zero and $c$ - is the ultimate velocity of propagation of
all types of interaction in nature do not have the real basis.

Really, from general solution of the non-stationary kinetic equation for the
photons (8), by the dividing the exponent to the series near the point $%
\stackrel{\rightarrow }{u}\stackrel{\rightarrow }{q}/\hbar \omega _q=1$ we
have

$$
N(q,t)=\left\{ N_o(q,T)+\frac{\beta N(q,T_i)}{\gamma _q}\right\} \left\{
1+\gamma _qt+\frac 12(\gamma _qt)^2+...\right\} -\frac{\beta N(q,T_i)}{%
\gamma _q}= 
$$

\begin{equation}
\label{19}=\left\{ N_o(q,T)(1+\gamma _qt)+N(q,T_i)\beta t\right\} +\frac
12\left\{ N_o(q,T)(\gamma _qt)2+\beta \gamma _qN(q,T_i)t^2\right\} 
\end{equation}
In the point $\gamma _q=0$ we have

\begin{equation}
\label{20}N(q,0)=\lim _{\gamma _q\rightarrow 0}N(q,t)=N_o(q,t)+N(q,T_i)\beta
t 
\end{equation}

As it is shown from this expression at the point $\stackrel{\rightarrow }{u}%
\stackrel{\rightarrow }{q}/\hbar \omega _q=1$, i.e. at the point $u=c$ the
distribution function of photons is non-stationary and grows linearly by the
time. What about of the singularity it is abbreviated clearly!

Since the Einstein theory is a stationary one it did not applied to the
non-stationary conditions, namely to the region of the drift velocities v =
c or u = c. For the drift velocities $v\geq c$ or $u\geq c$ the theory must
be non-stationary. The Einstein theory is a one mode theory and that is why
must be received from the many particle (or many mode) theory by the
limiting transition to the one mode case. The effect of connection charged
carriers and photons as a result of their mutual drag is a common
phenomenon. It is a reaction of the system on action of external fields for
the conservation the stationary state of system (the analog of
''self-conservation'' in biology). Thus we are found the ''dressing''
mechanism of charge carriers used earlier in quantum electrodynamics.

\section{Conclusions}

As a result of analysis the phenomena of connected system of charge carriers
and photons by the using of equations (4), (7) - (10) and (15) we receive
the following conclusions:

1. As it seems from non-stationary distribution function of photons the
Lorenz-Einstein theory corresponds to space uniform (equilibrium) case and
must satisfy the stationary condition $v<c$ ! The case of $v=c$ is not
included in their theory. In the point $u=c$ (i.e. $\gamma _q=0$) the
distribution function of photons (18) is non-stationary do not content any
singularity and grows linearly by the time.

For this reason the conclusions of Einstein theory that the rest mass of
photons is equal to zero and $c$ - is the ultimate velocity of propagation
of all types of interactions in nature do not have the real basis.

2. Since the Einstein theory was a one mode theory and that is why must be
received from the many body (mode) theory by the limiting transition to the
one mode case at $v<c$. For the $v>c$ or ($u>c$) the theory must be
nonstationary. As we say the mutual drag lead to the formation the
''quasi-particles'' - the electron or positron ''dressed by photons''. The
average energy in stationary state $u<c$ for the one mode case is:

$$
\left\langle \epsilon \right\rangle =\left\langle \hbar \omega \right\rangle
\left\langle N_i(q,T_i)\right\rangle =\frac{T_i\left\langle N(\omega
,T_i)\right\rangle }{1-u^2/c^2}= 
$$

\begin{equation}
\label{21}=\frac{T\left\langle N(\omega ,T_i)\right\rangle }{1-u^2/c^2}\frac{%
T_i}T=\frac{M_oc^2}{1-u^2/c^2}\left( \frac{T_i}T\right) ^2 
\end{equation}

The mass of the heated photons for one mode:

$$
M_i=\frac{M_o}{1-u^2/c^2}\left( \frac{T_i}T\right) ^2 
$$

The mass of one heated photon:

$$
m=\frac{M_i}{\left\langle N\left( \omega ,T_i\right) \right\rangle }=m_o 
\frac{T_i}T\left( 1-\frac{u^2}{c^2}\right) ^{-1}\text{ (21')} 
$$

At the absence of heating

$$
\left\langle \epsilon \right\rangle =Mc^2=\left\langle N_o\left( \omega
,T\right) \right\rangle \frac T{1-u^2/c^2}=M_oc^2\frac T{1-u^2/c^2}= 
$$

\begin{equation}
\label{22}=m_oc^2\frac{\left\langle N_o(\omega ,T)\right\rangle }{1-u^2/c^2} 
\text{; }m=\frac{m_o}{1-u^2/c^2}\text{,} 
\end{equation}
$m_o=M_o/\left\langle N_o(\omega ,T)\right\rangle =T/c^2$ is the rest mass
of photon, i.e. the mass of photon in frame of reference which drifted
jointly with photons at the temperature $T$, $\left\langle N\right\rangle $
is the concentration of photons for one mode.

3. As it seems from (19) at high electric and magnetic fields for the drift
velocities $u<c$ the energy (or the mass) of photons for one mode grows as a
result of the mutual drag, as well as the heating of the carriers and
photons.

4. As it seems from (4)-(10) at the external electric and magnetic fields
the relativistic factor enters to the expressions of the distribution
function and other physical quantities in first order in the form $\left(
1-u^2/c^2\right) ^{-1}$, instead the $\left( 1-v^2/c^2\right) ^{-1/2}$ in
Einstein theory. This is together with the conclusions 1 and 2 is solve the
main problem of super- luminal particles named -tachyons, because of in our
theory the imaginaryty of mass of tachyons is liquidated.

5. There is the opinion that the original conception about tachyons, as
individual particles such as the electrons, protons and etc. is not correct
and the tachyons in such understanding is absent [12]. {\bf Our
investigations shows that the ordinary particles such as the electrons,
positrons and also photons may stand a super-luminal in high external fields
under the conditions of mutual drag}. Also there are the opinion that the
tachyons as an elementary excitations (quasi-particles) have the wide -
spread in complex systems which is lose the stability and made the phase
transition to the stabile state [ibid] . In origin the tachyons in general
was considered only in amplifying mediums [13-16]. {\bf As it seems from our
investigations factually for the drift velocities more than the velocity of
light the super-luminal particles are generate or are amplify the
electromagnetic waves (photons) and the super-luminal particles are placed
on the regime of generation or amplification independently from on type of
medium (see also [17]). }As it will be shown in a special report in general
all elementary excitations including so- called '' elementary particles ''
are the quasi-particles.

6. In the point of $u=c$ the angle $\alpha $ between the $\stackrel{%
\rightarrow }{u}$ and $\stackrel{\rightarrow }{q}$ is equal to $\pi $$/2$
and we have the condition when the electromagnetic wave becomes a free and
is emitted. Thus the point of $u=c$ is the point of transition of system
from absorption regime to the regime of emission of electromagnetic waves
(photons).

7. It is shown that the relativistic expression for the deceleration of time
really taking place for the relaxation time of carriers on photons, for the
life-time of carriers and also for the period of electromagnetic
oscillations. Factually, in dynamics and electrodynamics the time is enter
as a parameter, but not as a free coordinate and for this reason the
Einstein relation is impossible to apply to the time.

Since $\tau ^{-1}\symbol{126} N_i(q,T_i)$ we have $\tau _i\approx \tau
_o\left( 1-u^2/c^2\right) \left( T/T_i\right) $; $l_i=u\tau _i=l_o\left(
1-u^2/c^2\right) \left( T/T_i\right) $. If $T_i=T$ we have $\tau _i\approx
\tau _o\left( 1-u^2/c^2\right) $; $l=l_o\left( 1-u^2/c^2\right) $.

8. It is shown that the so-called ''velocity of light'' in vacuum $-c$, as
well as the velocity of sound for phonons, is an averaged velocity of
photons at the ground state.

9. As it seems from (4), (7), (8) and (10) at the point $u=c$ the
distribution function of photons is an isotropic one and the anisotropy part
of photons distribution function at this point is equal to zero. It means
that the ground and the stationary states of electromagnetic field is
spherically symmetric (this question will be discussed in a special report)
.At the point $u=c$ the stimulated absorption and emission are equal to each
other and there are only spontaneous emission of photons.

10. It is shown that the demand for invariance of the Maxwell equations or
the laws of physics in all inertial frames of reference is not correct. It
is equivalent to the demand of invariance the lows of physics in the uniform
and non-uniform spaces or to the demand of equivalency the lows of physics
in cases of the presence and absence of external field (force).

11. It is shown that the presence of the second inertial frame of reference
moving with the constant drift velocity relatively to first one may be a
result of the presence of space non-uniformity or the non-uniform external
field.

Really in the uniform space because of the equivalency of all points of
space it is impossible to get simultaneously two, or more inertial frame of
reference with the different constant drift velocity relatively to each
other. Because it will lead to the violation of the space uniformity, i.e.
to the change of the distance $r_{12}$ between the initial points of that
frames of reference ($r_{12}\neq const.$). In the uniform space for the
conservation of the space uniformity during the motion it is necessary a
motion of all points of the space with the same constant velocity (in the
absence of the external field), or with the constant acceleration (in the
presence of the uniform force). For the both cases the frames of reference,
connected with the different points of the space, do not have the motion
relatively to each other without the violation of the space uniformity or
the uniformity of the external force. Since all points of the spaces in both
cases are placed on the same conditions and are equivalent. In the second
case for the reason that the force in the Newton's second law is external
one the system (or the space) must be opened. To choose two frames of
reference moving with the constant drift velocity relatively to each other,
it is necessary to have the space, which consists of two half-spaces. All
points of first half-space are in rest or have the motion with the constant
velocity v and all points of second half-space are move with the constant
acceleration. The first of them is inertial, but the second accelerates and
for this reason the demand of equivalency of laws of physics in that two
frames of reference is equivalent to the demand of the equivalency of first
and second Newton's laws. This demand is absurd, of course!

12. As it seems from (1) - (4) the demand of the equivalency of laws of
physics in the all inertial frames of reference for the photons is
equivalent to the demand about equivalency the Plank' s equilibrium
distribution function to the drifted Plank's distribution function.

13. It is shown that at the external electric fields $E<H$ the drift
velocity of carriers $u<c$ and the energy received from external field is
accumulated by the connection of carriers with photons, as a result of the
mutual drag (as a result of the construction of structure by the mechanism
of '' dressing of carriers with photons '') and stationary state is
conserved. In this region of the drift velocities the absorption is more
than the emission. Under the conditions $E>H$ i.e. at the drift velocities $%
u>c$ the generation and amplification is dominate and there are the
exponential grown the number of photons by the time. {\bf The violation of
the stationary state is begins from the point $u=c$ and from this point is
begins the transition of the carriers'' dressed by photons '' to the
following stationary state by the reactive emission of photons}. In the
region of drift velocities $u>c$ the anisotropy part of the photons
distribution function is much more than the isotropic one. {\bf Thus in our
investigation the mechanism of the transition of particles to the following
stationary state is obtained.}

14. It is shown that for the drifted velocities $u>c$ so - called energy (or
mass) for one mode are:

$$
\left\langle \epsilon \right\rangle =\left( \frac{T_i}T\right) ^2\cdot \frac{%
T\left\langle N(q,T)\right\rangle }{u/c-1}\left\{ \left[ \exp \left( \gamma
_qt\right) -1\right] +\frac T{T_i}\exp \left( \gamma _qt\right) \right\} = 
$$
\begin{equation}
\label{23}=\frac{M_oc^2}{u/c-1}\left( \frac{T_i}T\right) ^2\left\{ \left[
\exp \left( \gamma _qt\right) -1\right] +\frac T{T_i}\exp \left( \gamma
_qt\right) \right\} 
\end{equation}

or the mass for one mode:

$$
M=\frac{M_o}{1-u/c}\left\{ \left[ \exp \left( \gamma _qt\right) -1\right] + 
\frac{m_o}{m_i}\exp \left( \gamma _qt\right) \right\} \text{ (23')} 
$$

In the absence of heating

\begin{equation}
\label{24}M=\frac{M_o}{u/c-1}\left[ 2\exp \left( \gamma _qt\right) -1\right]
\approx \frac{2M_o}{u/c-1}\exp \left( \gamma _qt\right) 
\end{equation}
As it seems from this equations the mass of photons for one mode in this
region of drift velocities is $u>c$ grows by the time exponentially.

The case considered by Einstein may be correspond to the case when from two
choosing frame of reference the first was connected with the photons and
drifted together with them and the second was connected with the charged
carriers (electrons or positrons) drifted with constant velocity relatively
to first one. Here the photons (electromagnetic field) is plays the role of
media and the charged carriers drifted relatively to that media. Thus the
system must consists of three subsystems. Factually, in electrodynamics the
space (or the system) is consists from three half - space (or subsystems):
the half -system of negatively charged carriers, half-space of positively
charged carriers and the half-space (or space) of photons. For this reason
the electrodynamics space is non-uniform initially, because of the point of
space where is placed of negatively charge carrier do not equivalent to the
point of space where is placed the positively charged carrier and both do
not equivalent to the point where is placed the photon. The presence of the
two type of charge lead to the presence of so-called Lorentz force. The
condition of stationary of state is the equality of this force to zero $F=0$%
! This condition corresponds to the annihilation of charge carriers with
production of photons, i.e. production of free electromagnetic field without
charges. It means that the space of photons (i.e. free electromagnetic
field) can decay to the two half-spaces: the spaces of the negative and
positive charges and also two half-spaces of negative and positive charges
can product the space of photons (the electromagnetic space or media).

As it seems from our investigations considered by Lorentz and Einstein case
corresponds to the case of the presence of weak external field when the
heating of carriers and photons are absent and there are only their mutual
drag. Also they was consideredthe case of one type of charged carriers. In
the presence of mutual drag of the electrons and photons the distribution
function of photons $N(q)$ has the form of the displaced Plank's function
with the constant drift velocity and as a result with the renormalized
frequency of emission $\omega _{em}=\omega _q^{*}=\omega _q-\stackrel{%
\rightarrow }{u}\stackrel{\rightarrow }{q}/\hbar $. For the drift velocities 
$u<c$ is decrease with the increasing of drift velocity $u$ because of the
stationary function of photons has the form (18) or (10).

Thus in the second frame of reference charge carrier is emitted or absorbed
photon with the frequency $\omega _{em}^{*}=\omega _{abs.}\left[ 1-\stackrel{%
\rightarrow }{u}\stackrel{\rightarrow }{q}/\hbar \omega _q\right] $. In the
one mode case the observed frequency $\omega _{abs}=\stackrel{\_}{\omega }%
=T_i/\hbar $. In the other word

$$
\omega _{abs.}=\omega _{em.}\left( 1-\frac uc\cos \alpha \right) ^{-1}\text{
or }\lambda _{abs.}=\lambda _{em.}\left( 1-\frac uc\cos \alpha \right) 
$$
For the case when the both frames of reference assumed to be inertial one,
i.e. to move along the one line (along the $x$-axis) cos$\alpha =1$ and we
have

$$
\omega _{abs.}=\omega _{em.}\left( 1-\frac uc\right) ^{-1}\text{ or }\lambda
_{abs.}=\lambda _{em.}\left( 1-\frac uc\right) 
$$
As it seems from this equation the Doppler effect is also a result of the
mutual drag of carriers and photons. Factually the source of emission
(charge carrier) drifts relatively to frame of reference connected with the
photon (observer) with the drift velocity $\stackrel{\rightarrow }{u}$. By
the increasing of drift velocity $\stackrel{\rightarrow }{u}$ the distance
between the source and the detector are increased too and as a result the
observed frequency of photons is increased or the wavelength is decreased.
In the opposite case the wavelength is increased. In the region of $u<c$ the
source (the charge carrier) moves slowly than the detector (photon) and as a
result the distance between them is decreased and the wavelength of observed
photons (light) is increased. Thus the mutual drag of electrons and photons
in the region $u<c$ lead to decreasing the frequency of emission or
absorption (see ''Low frequency cyclotron resonance for the phonons ''
[2,5,9]). It means that the frequency of observed light (photons) is
increased.

15. As it seems from the present consideration the so-called relativistic
phenomena and the Doppler effect is a same ones and are a result of the
mutual drag of interacting system of carriers and photons at external field.
The case was considered by Lorentzand Einstein corresponds to the case, when
the charge carriers are drifted at the electromagnetic field under the
conditions of the mutual drag of carriers and photons in the absence of
their heating by the field.

Distinction between the results of the theory of relativity and the Doppler
Effect is connected with them, that the Doppler Effect is dealing with the
total stationary distribution function but no only it's isotropic path as in
theory of relativity. In the other words the Doppler Effect is received as a
result of taking into account the violation of the $T$-symmetry at the
external field.

\section{Reference}

1. Gasymov T. M., 1976, Dokl. Acad. Nauk. Azerb. Ser.Fiz.-mat.Nauk, 32(9),
19-22.

2. Gasymov T. M., 1977, Nekotorye Voprosy Eksp.Teor.Fis.,Baku, Elm, p.p.
3-27.

3. Gasymov T . M., 1990, Nekotorye Voprosy Teor.Fis.,Baku, Elm, p.p. 139-150.

4. Gasymov T. M..,Katanov A.A.1990, J. Phys. Condens. Matter 2,
p.p.1977-1991.

5. Gasymov T. M., 1976, Dokl. Acad. Nauk. Azerb. Ser.Fiz.-mat.Nauk, 32 (6),
15-17.

6. Gasymov T .M., Katanov A. A.,1974, Izvest. Akad. Nauk Azerb. No 4, p.p.
44-53.

7. Gasymov T. M. 1976. Dokl. Acad. Nauk Azerb. SSR, 32, No 6, p.p. 15 - 17.

8. Gasymov T.M., Katanov A.A.,1974, Phys.Stat.Sol. (b), 64, p.p. 557-566.

9. Bass F.G. 1966, JETF, pisma, 3, 357.

10. Korenblit I.Y. 1967, JETF, 53, 300.

11. Gasymov T. M., Granovsky M.Y., 1976, Izvest. Acad. Nauk Azerb. No1,
p.p.55-64.

12. Andreev A.U., Kirjnis D.A. 1996, Uspechi Phys. Nauk , 166, No 10, p.p.
1135-1140. [U1]

13. Bassov N.G. at all.1965. DAN SSR,165, p.p.58; 1966. JETF,50,p.p.23.

14. Raymond Y. Chiao, Kozhekin A.E. and Kurizki G. 1966. Phys.Rev.Lett.77,
No7, p.p.1254-1257.

15. Orayevsky A.N.1998.Uspechi Phys. Nauk,168, No 12, p.p.1312-1321.

16. Barashenkov V.S.1974.Uspechi Phys. Nauk,114,No 1, p.p.133-149.

17. Tofig M. Gassym. 1999.Circular of Shamakhy Astrophys. Observ. No 96,
p.p.33-35.

\end{document}